# Selecting optimal subgroups for treatment using many covariates

Tyler J. VanderWeele[1]
Alex R. Luedtke[2]
Mark J. van der Laan[3]
Ronald C. Kessler[4]

1 Departments of Epidemiology and Biostatistics, Harvard T.H. Chan School of Public Health, Boston, Massachusetts

2 Fred Hutchinson Cancer Research Center, Seattle, Washington

3 Department of Biostatistics, University of California, Berkeley, California

4 Department of Health Care Policy, Harvard Medical School, Boston, Massachusetts

**Abstract**

We consider the problem of selecting the optimal subgroup to treat when data on covariates is available from a randomized trial or observational study. We distinguish between four different settings including (i) treatment selection when resources are constrained, (ii) treatment selection when resources are not constrained, (iii) treatment selection in the presence of side effects and costs, and (iv) treatment selection to maximize effect heterogeneity. We show that, in each of these cases, the optimal treatment selection rule involves treating those for whom the predicted mean difference in outcomes comparing those with versus without treatment, conditional on covariates, exceeds a certain threshold. The threshold varies across these four scenarios but the form of the optimal treatment selection rule does not. The results suggest a move away from traditional subgroup analysis for personalized medicine. New randomized trial designs are proposed so as to implement and make use of optimal treatment selection rules in health care practice.

# Introduction

Biomedical researchers and social scientists are often interested in identifying the subgroups that would benefit most from a particular treatment or intervention. In randomized trials, subgroup analyses are often used to compare the effect of treatment across subgroups defined by various pre-treatment covariates (Yusuf et al., 1991; Assmann et al., 2000; Pocock et al, 2002; Rothwell, 2005; Lagakos, 2006; Wang et al., 2007). Such analyses can help give insight into whether a treatment might be more effective for men versus women, or for younger versus older persons, or for any other characteristic or variable defined prior to receipt of treatment. These types of analyses are relevant if the effect of treatment might vary across individuals in a population, a phenomenon often referred to as "effect heterogeneity." Such analyses can be useful in deciding who to treat, or who to treat first, if resources are limited. They can also be useful when deciding which of two treatments to give to whom.

While well-established methodology has been used for decades to carry out such subgroup analyses across strata defined by a single covariate (Rothman et al., 1980; Rothman, 1986; Yusuf et al., 1991; Hosemer and Lemeshow, 1992; Li and Chambless, 2007; VanderWeele and Knol, 2014), in actual practice it would be more desirable to make use of data on numerous covariates. Viewed from the individual perspective, we are interested in knowing how to best choose the appropriate treatment for an individual with a particular set of characteristics. This task is sometimes now described as "personalized medicine" or "precision medicine." It is the optimal selection of treatment for the individual (Murphy, 2003; Robins, 2004; Chakraborty and Moodie, 2013). However, viewed from the perspective of a population, if we optimize the treatment for each individual, we are also optimizing the outcomes for the population and are thus interested in which subgroups to give which treatment in order to maximize the outcomes within a population of interest, possibly subject to resource constraints.

To make progress with multiple covariates for this task, it is not uncommon in the biomedical or social sciences to form a "prognostic score" (Rothwell, 1995; Hayward et al., 2006; Kent and Hayward, 2007; Pocock and Lubsen, 2008; Abadie et al., 2013). In a randomized trial with treatment and control, this prognostic score is defined as the predicted value of the outcome, conditional on an individual's covariates, if that person were not given treatment. The prognostic score is often obtained by first fitting a regression model of the outcome on the pre-treatment covariates among the control arm of the randomized trial. Using the estimates of the regression parameters of this model, one can then obtain predicted outcomes under the absence of treatment for each individual in the study to give the prognostic score. The prognostic score itself is then typically taken as the variable by which subgroups are formed. An analyst might for example, subsequently analyze the data within tertiles, quartiles, or quintiles of the prognostic score. If those with low prognostic scores would benefit most from treatment then this might be the group for which it would be best to target treatment. This approach is used with some frequency in the biomedical and social sciences (Kent et al., 2002; Fox et al., 2005; Rothwell et al., 2005; Abadie et al., 2013; Pane et al., 2014). It is sometimes also referred to as "risk stratification" (Kent and Hayward, 2007) or "endogenous stratification" (Abadie et al., 2013). While such procedures theoretically are effective with very large sample sizes, recent evidence suggests that in most practical settings, even with thousands of study participants (Abadie et al., 2013), biases from this sort of approach result from overfitting if the same data are used to form the prognostic score and to run the subgroup analyses (Peck, 2003; Hansen, 2008; Abadie et al., 2013). Various techniques, using cross-validiation, have been proposed to address these biases (Abadie et al., 2013).

However, a more fundamental problem with the approach is that even if such biases were absent, using the prognostic score or individual-covariate subgroup analysis, does not in fact identify the optimal treatment allocation rule. There are better ways to use the covariate data available to optimize an individual's outcome and the mean outcomes for the population. A growing literature has begun to explore statistical approaches for more effective treatment selection rules (Cai et al., 2011; Zhao et al., 2013; Luedtke and van der Laan, 2015, 2016ab). In fact, what such an optimal rule depends subtly on precisely what question the analysis is intended to address.

In this paper we will present four settings in which optimal subgroup selection is of interest. We will describe these settings and the optimal treatment rule in each. We will discuss how the approaches in this paper relate to what is typically done in practice and how might be best to proceed in subsequent research when selecting optimal subgroups for treatment is of interest. New randomized trial designs are further proposed so as to implement and make use of optimal treatment selection rules in practice.

**Notation**

We will let A denote a treatment or intervention under study. We will assume that receipt of treatment has been randomized with probability 1/2 but we will comment later in the paper on how the methodology described here is also potentially applicable to observational studies. We will let Y denote an outcome of interest. Finally we will let C denote a set of pre-treatment covariates that are available for each individual in the study. We will let $Y_1$ denote the potential outcome (Rubin, 1974) that would have occurred for each individual if they had received treatment and we will let $Y_0$ denote the potential outcome that would have occurred under control. We only get to observe one of $Y_1$ and $Y_0$: we observe $Y_1$ for those who actually received treatment and $Y_0$ for those who were actually in the control arm. We do not in general know the potential outcome if an individual had been in the other arm of the trial.

In what follows, the task of treatment selection will essentially be to partition the population into two groups, which we will call "T" and "S", those receiving the treatment, and those not receiving the treatment, respectively. The goal will be, in each setting, to decide on how to partition the population into those who do versus do not receive treatment in order to maximize mean outcomes. We will refer to this partition of individuals who do and do not receive treatment as the optimal treatment rule.

We will, for simplicity, here assume that treatment A is binary with 1 denoting treatment and 0 denoting control. However, the ideas that are developed below are also applicable if we are comparing two different treatments so that A=1 denotes one treatment and A=0 denotes another. Although we will generally use of "treatment" and "control", the same methods and ideas described below are applicable also in the setting of comparing two treatments with "selecting who gets treatment" simply interpreted as "selecting who gets the first treatment" and "control" interpreted as "those receiving the second treatment." In the discussion section we will also comment on how the ideas potentially extend to settings when more than two treatments are being considered.

In what follows we will provide an overview of the relevant concepts and methods. We will state results that are precise under some technical conditions. More formal statements and proofs are given in the Online Appendix and elsewhere (Luetke and van der Laan, 2015, 2016ab).

**Four Questions Relevant to Optimal Subgroup Selection**

We will consider four settings that may be of interest in selecting optimal subgroups for treatment. Stated intuitively, these settings are:

1. Who do we treat if resources are limited so that we can only treat q% of the population?
2. Who do we treat if resources are not limited so that we could potentially treat everyone and are simply deciding who would benefit from treatment?
3. Who do we treat if resources are not limited, but are subject to costs or side effects?
4. How do we select subgroups to maximize the "effect heterogeneity" across subgroups?

We will address each question in turn.

*Setting 1. Subgroup selection under resource constraints*

First let us suppose that due to some form of resource constraints (e.g. costs, doses available, etc.), we are only able to treat at most q% of the population. We have data from a randomized trial of treatment A where we have collected outcome Y and pretreatment covariates C. We want to use the covariates C, and the outcome data from our randomized trial to determine a treatment rule in order to partition the population into those that we should treat so as to maximize the expected outcome for the population, subject to the constraint that we can only treat q% of the population. Once we decide on these two sets, T, the treated, and S, the untreated, then the expected outcome for the population under this treatment rule is:

$$\frac{q}{100}E[Y|A=1,T] + (1-\frac{q}{100})E[Y|A=0,S] \qquad (1)$$

In other words, for q% of the population we get the average outcome under treatment for the subgroup T that we selected for treatment and for (100-q)% of the population we get the average outcome under control for the subgroup S that we selected not to receive treatment.

It is shown in the eAppendix that if we knew the potential outcomes, $Y_1$ and $Y_0$, for each individual in the population then the optimal treatment rule to maximize the expected outcome for the population would simply be to treat those for whom {$Y_1$-$Y_0$>k} where k is determined so that exactly q% are treated. In other words, if we knew the potential outcomes for each individual, so that we knew the actual effect, $Y_1$-$Y_0$, of treatment for each individual, we would simply treat the q% for which the effect of treatment itself was largest. In actual fact however, we do not know both potential outcomes for every individual in the population. We only have our randomized trial data, our outcomes Y, and our covariates C. So we want to use C to partition individuals into those who we do or do not treat to maximize outcomes. It is again shown in the eAppendix that to maximize outcomes, using covariates C, the optimal treatment rule is to treat those with covariate values c such that

$$\{E[Y|A=1,C=c]-E[Y|A=0,C=c]>k\}$$

where the cut-off k is again determined so that exactly q% are treated. In other words, the optimal treatment rule is to treat the q% with the highest expected treatment effect conditional on their covariates. The expected treatment effect for each individual conditional on their covariates is something that can be estimated from the data in a randomized trial and thus this treatment rule can be implemented in practice. We could, for example, fit regression models for the expected outcome under treatment E[Y|A=1,C=c] and under control E[Y|A=0,C=c], (or, more directly, their difference) conditional on covariates, to obtain estimates. In the next section, we will discuss some statistical issues relevant to implementing this in practice. However, again, it can be shown, that the best we can do in terms of maximizing outcomes for the population using just the covariates C is to treat those with the highest expected treatment effect, E[Y|A=1,C=c]-E[Y|A=0,C=c], conditional on their covariates. With this treatment rule the expected outcome for the population is again then qE[Y|A=1,T] + (1-q)E[Y|A=0,S].

The expected outcome under the treatment rule will not be as high as we could have obtained had we known both potential outcomes for all individuals, but again this is the best we can do with the measured covariates C. We could compare the expected outcome (1) under the treatment rule to what we would obtain if we simply randomly selected q% of the population for treatment, in which case we would have an expected outcome of:

$$\frac{q}{100}E[Y|A=1] + (1-\frac{q}{100})E[Y|A=0] \qquad (2)$$

How much better we do under the treatment rule using the covariates C will depend in part on how predictive the measured covariates are of the association between treatment and the outcome of interest, and also how well we statistically model the expected outcomes $E[Y|A=1,C=c]$ and $E[Y|A=0,C=c]$, or, more directly, their difference. We could compare the expected population outcomes in (1) under different estimates of the optimal treatment rule using different modeling techniques. Again, in the next section we will consider issues of statistical modeling. Intuitively, how well we improve on the outcomes by selecting subgroups for treatment using covariates C, instead of randomly allocating treatment, will effectively depend on how well we can use the covariates C and statistical modeling to predict the potential outcomes. i.e. how well we estimate the true $E[Y|A=1,C]-E[Y|A=0,C]$.

*Setting 2. Subgroup selection under unconstrained resources*

We will now turn to a different setting in which resources are not constrained so that we could potentially treat anyone who might benefit from treatment. Once again our objective is to determine the treatment rule that partitions individuals into two sets: T, those who do receive treatment, and S, those who do not; so as to maximize the average outcome for the population, which is then:

$$E[Y|A=1,T]P(T) + E[Y|A=0,S]P(S) \qquad (3)$$

It is shown in the eAppendix that if we knew the potential outcomes, $Y_1$ and $Y_0$, for each individual in the population then the optimal treatment rule to maximize the expected outcome for the population would simply be treat those for whom $\{Y_1-Y_0>0\}$. In other words, if we knew the potential outcomes for each individual, we would simply treat those for whom the effect of treatment itself was positive. This is, of course, relatively intuitive. In actual fact, we do not, of course, know both potential outcomes for every individual; we only have our covariates C. With covariates C, it shown in the eAppendix that to maximize outcomes, using covariates C, the optimal treatment rule is to treat those with covariate values c such that

$$\{E[Y|A=1,C=c]-E[Y|A=0,C=c]>0\} \qquad (4)$$

In other words, we treat those who have, conditional on their covariates, a positive expected treatment effect. We can again estimate this from the data from our randomized trial and statistical aspects of such estimation are again briefly noted in the next section. Under this treatment rule, the expected outcome will simply be E[Y|A=1,T]P(T) + E[Y|A=0,S]P(S). We could compare this expected outcome under the optimal treatment rule, to the expected outcome if we treated everyone in the population, E[Y|A=1], or if we treated no one, E[Y|A=0]. The extent to which we can maximize the outcome using covariates C will again depend in part on how predictive the covariates C are of the outcome of interest, and also how well we statistically model the expected outcomes E[Y|A=1,C=c] and E[Y|A=0,C=c], i.e. on how well we can use the covariates C and statistical modeling to predict the potential outcomes.

An interesting feature of this second setting of unconstrained optimal treatment selection is that the tasks of individual decision-making and maximizing population outcomes in fact coincide. The approach to maximize population outcomes is simply to assign treatment to anyone who would benefit from it. The perspectives of the individual and the policy-maker coincide. This was not the case in the first setting wherein an individual might have a positive expected treatment effect and therefore, from an individual perspective, have expected benefit from treatment, whereas a policy-maker, to maximize population outcomes, might choose not to treat that individual because others have higher expected treatment effects and resources are limited.

*Setting 3. Subgroup selection under costs and side-effects*

Now let us turn to a setting in which resources are not constrained so that we could once again, in principle, treat everyone but now suppose the treatment itself has a cost that we want to take into account, and/or has side-effects that we want to weigh against the potentially beneficial effects on our outcome of interest Y. Because of costs or side effects we might, for example, only want to treat those with treatment effects larger than some level $\delta$. Or more generally, that level might depend on a person's covariates c so that we only want to treat those with treatment effect greater than some level $\delta(c)$. The optimal rule (see eAppendix) if we knew both potential outcomes for all individual would then simply be to treat those with $\{Y_1-Y_0>\delta(c)\}$ and the optimal rule with the actual trial data and measured covariates C would be to treat those with $\{E[Y|A=1,C=c]-E[Y|A=0,C=c]> \delta(c)\}$. Once again, how well we could optimize outcomes would depend on how predictive the covariate C were of the association between treatment and outcome.

*Setting 4. Maximizing effect heterogeneity*

When one reads through the subgroup analyses of many randomized trials, in which subgroup analyses are undertaken one covariate at a time, it often seems that the goal is to find a covariate, often dichotomous or dichotomized, such that the effect heterogeneity across subgroups defined by the covariate is as large as possible.

When the effect estimate in one subgroup is much larger than that of the other, then the subgroup analysis is considered a success and that covariate defining the subgroups is subsequently considered important. In fact, we could carry out a similar exercise using data on multiple covariates. In this case, we would want to use covariates C to partition the population into two subsets, T and S, such that the effect in the subgroup T, E[Y|A=1,T]-E[Y|A=0,T], was much larger than the effect in subgroup S, E[Y|A=1,S]-E[Y|A=0,S]. In other words, we would want to maximize effect heterogeneity by maximizing the difference between the effects in these two subgroups:

E[Y|A=1,T]-E[Y|A=0,T]  -  {E[Y|A=1,S]-E[Y|A=0,S]}

This is in some sense a generalization of what seems to be the traditional subgroup task but extended to multiple covariates simultaneously. It is shown in the eAppendix that once again the solution to this maximization takes the form of selecting T to be those with an actual treatment effect, $Y_1-Y_0$ (if the potential outcomes were known), or expected treatment effect conditional covariates C, E[Y|A=1,C=c]-E[Y|A=0,C=c], above some threshold k', where k' can be determined numerically as described in the eAppendix. But once again, it is the expected conditional treatment effect, E[Y|A=1,C=c]-E[Y|A=0,C=c], that is utilized in the criterion by which treatment decisions are to be made in this setting as well. Note, however, that although this treatment rule maximizes effect heterogeneity, the average outcome under this treatment rule will generally be worse than that selected by the treatment rule that maximizes the outcome itself as in setting 2. It is thus not clear that this treatment rule that maximizes effect heterogeneity is of particular use in decision-making, unlike those in contexts 1, 2 and 3 above. We will return to this point in the discussion.

*Extensions to Observational Studies*

Our discussion thus far has been within the context of a randomized trial. However, as discussed further in the eAppendix, all of the discussion above pertains also to optimal subgroup selection and treatment decisions from data arising from an observational study as well, provided that the covariates C suffice to control for confounding of the effect of treatment A on outcome Y, though the formulae for the optimized outcome need to be modified (see eAppendix). With data from observational studies, an additional context that may be of interest is if, in data from the study, there are available covariates C that suffice to control for confounding for the effect of treatment A on outcomes Y, but if, when treatment decisions are made subsequently, only data on some subset W of the covariates C will be available. Methodology for this setting has been developed and is described elsewhere (Luedtke and van der Laan, 2015; 2016ab).

**Statistical Analysis**

In the eAppendix we describe methods and formal statistical inference for estimating the optimal treatment rule and the outcome under it, as well as software to do so. While there are many ways to go about estimation, the methods described in the eAppendix flexibly model the difference in observed outcomes across treatment groups conditional on the covariates and use an ensemble technique called "super-learner" (van der Laan et al., 2007) that considers numerous different possible models or algorithms for the conditional outcome differences and then weights these according to their mean square error predictive value using cross-validation. Statistical inference for the optimal treatment rule and for the outcome under it is challenging because the same data are being used to estimate the treatment rule and the expected outcome under it. Sample-splitting can potentially be used but is not efficient, and averaging across split samples does not yield valid inference (van der Laan and Luetke, 2014). The eAppendix describes a cross-validated targeted minimum loss-based approach to estimate the optimal treatment rule and the outcome under it. While the approach described in the eAppendix has some desirable theoretical properties, considerable work remains to be done in assessing the sample sizes that are needed for these techniques to be useful and how the various methods that have been proposed in the literature compare to one another in actual practice. While the theoretical methodological development has come a long way in the past decade, much remains to be learned about the application of these methods. Our focus in this paper is on conceptual foundations for the various methods that have been proposed, and the implications of these conceptual foundations for interaction and subgroup analyses within epidemiology, the topic to which we now turn.

**Discussion**

In this paper we have shown that under a wide range of different goals and settings, including making treatment decisions with or without resource constraints, and with or without side effects, or even when trying to maximize effect heterogeneity, the correct approach to finding the optimal treatment rule is to estimate expected treatment effects for each individual conditional on the covariates. In each of the settings described above, the optimal treatment rule involved treating those above some threshold of the conditional expected treatment effect. The threshold differed according to whether there were or were not resource constraints, or whether there were or were not costs or side effects, or whether we wanted to maximize effect heterogeneity, but the form of the treatment rule did not vary across these contexts. In each case, the form of the optimal treatment rule was simply to treat those with conditional expected treatment effects above a specific threshold. This has a number of important implications for the actual practice of subgroup analysis, treatment selection, precision medicine, and the modeling of interactions.

One fundamental insight from our discussion above is that for treatment selection and decisions, our discussion suggests a need to move away from subgroup analyses conducted one covariate at a time. The problems with this approach are numerous. First, subgroups may come into conflict: if subgroup analyses indicate that treatment A is better for women and treatment B is better for men, and also indicate that A is better for aged and B better for younger persons, and we want make treatment decisions for a younger woman, the subgroup analyses conflict. Second, the subgroup analyses often fail to answer the scientific question of interest. As they are typically carried out, they tend to be aimed at maximizing effect heterogeneity, whereas what is actually of interest is maximizing population outcomes or individual treatment decision making. The optimal treatment rule for maximizing population outcomes or individual treatment decision-making is not the same as for maximizing effect heterogeneity. Finally, compared to individual covariate subgroup analyses, we can in fact do better at maximizing mean outcomes by making simultaneous use of all covariates, rather than running analyses one covariate at a time. It is conceivable, of course, that the optimal treatment selection in some rare cases might involve only a single dichotomous covariate, or in some settings a single dichotomous covariate may constitute the decision to be made (e.g. resources are limited so we can only city 1 or city 2), but in general the optimal decision-making rule will make fuller use of covariate data.

The need to move away from one-covariate-at-a-time approaches in optimizing population outcomes or individual treatment decision-making is relevant not just to traditional subgroup analyses when we are looking at whether the treatment effect is larger in one group versus another, but this same point is also relevant to the analysis of so-called "qualitative" or "cross-over" interactions (Gail and Simon, 1985; Piantadosi and Gail, 1993; Pan and Wolfe, 1997; Silvapulle, 2001; Li and Chan, 2006), in which the treatment has a positive effect in one subgroup and a harmful effect in another. The analysis of such cross-over interactions is again often done one covariate at a time, but for the purposes of decision-making, it ought to be done using all available relevant covariate data. In actual fact, the methodology described in setting 2 above is doing precisely that.

A second important implication of the discussion in this paper is that, for optimizing population outcomes or individual treatment decision making, we should move away from the "prognostic score", that is often employed in both the biomedical and social sciences (Hayward et al., 2006; Kent and Hayward, 2007; Fox et al., 2005; Rothwell et al., 2005; Abadie et al., 2013; Pane et al., 2014). The practice of stratifying on prognostic scores in small- or medium- sized trials has numerous statistical problems with "overfitting" documented elsewhere (Abadie et al., 2013). At a more fundamental level, though, it gets the objective wrong because the patients at greatest risk of bad outcomes in the absence of treatment are not necessarily the same patients who will profit most from intervention. While stratifying the results of randomized trials using the predicted outcome under control can provide some insight into who might be considered to have greatest need for treatment, it is not the correct approach to optimize population outcomes or individual treatment decision-making. To optimize population outcomes or individual treatment decision-making, one stratifies, not by predicted outcome under control, but by the expected effect of treatment; that is, the difference between the predicted outcome under treatment and the predicted outcome under control, conditional on covariates. It is this stratification that gives one insight into optimal treatment decisions either with or without resource constraints.

A third important implication, related somewhat to the first, concerns the modeling of interactions. In reading the literature, one is often left with the impression that the principal goal of interaction analysis is to determine whether, in a given statistical model, a product term involving two variables is "statistically significant" or non-zero. Methodology to detect to "interactions" or non-zero product terms has become increasingly advanced (e.g. Moore et al., 2006; Green and Kern, 2012; Imai and Ratkovic, 2013; Berger et al., 2015). However, once again, if the purpose of the analysis is optimizing population outcomes or individual decision-making, the question as to whether a specific product term in a particular statistical model is present is, in fact, secondary. All that matters for the task of optimizing population outcomes or individual decision-making is having predictive covariates and having statistical models that give good predictions of expected outcomes conditional on those covariates. If the product terms help in a particular model, they can be included; if not, they can be omitted. In either case, though, their presence or absence is secondary to having a good predictive model so as to make optimal treatment decisions. Indeed using models both with and without product terms and, more generally, numerous models and machine learning algorithms, to generate predicted outcomes, and possibly ensemble methods to average over, or choose among them, as suggested above, is probably a preferable way to proceed.

It might be thought that subgroup analyses one-covariate-at-a-time or the analysis of individual product terms in statistical models may still be of interest for the purposes of understanding or explanation. While this may be true to some degree, it is important to clarify the goal of such understanding or the form of explanation that is in view. If what is thought to be of importance is to understand which covariates in fact are most relevant in decision-making (e.g. because it was thought undesirable to measure all of the covariates subsequently in treatment decision-making), then one could instead consider the result of optimal treatment rules on the maximized population outcome when only certain subsets of the covariates C are considered. On the other hand, if one simply wanted to assess which covariates in some sense seemed most "responsible" for the effect heterogeneity, one might instead still model the outcome with all covariates simultaneously, and then consider, for example, what a one-unit shift in any given covariate for all individuals would have in changing expected treatment effects. In linear models for the expected outcomes in each treatment arm, this would simply be the difference between the covariate coefficient in the model under treatment $E[Y|A=1,C=c]$ and the covariate coefficient in the model under control $E[Y|A=0,C=c]$. But the approach of considering a one-unit shift in a particular covariate across all individual could also be employed in non-linear models as well. Other metrics could also potentially be developed. Finally, sometimes analyses of interactions are undertaken for the purpose of understanding the joint effects of the treatment and a particular covariate, or to gain mechanistic insight (VanderWeele and Robins, 2007; VanderWeele, 2009, 2015). In this case, it may be appropriate to assess the joint effects of one covariate at a time, but in this case, if the *effect* of the covariate is in view, then confounding control must be made for the association between that covariate and the outcome (VanderWeele, 2009; VanderWeele and Knol, 2011; VanderWeele, 2015) and what additional variables are needed to control for such confounding will vary depending on which covariate is in view. This is no longer simply a question of effect heterogeneity but of *joint effects* (VanderWeele, 2009; VanderWeele and Knol, 2014). A model which includes all of the covariates C available will not in general be adequate to provide appropriate control in addressing this type of question if the covariates themselves affect one another.

Yet another argument that might be put forward for doing one-by-one subgroup analyses may involve trying to generate heuristics. A physician cannot remember the functional form of two conditional expectations but can remember that treatment A is better for women and treatment B is better for men. While such treatment heuristics can be of some value, they can, as already discussed above, come into conflict with one another. Moreover, the use of such heuristics in decision-making becomes even more complex when there are more than two potential options to choose among, which brings us to another topic of our discussion: extensions to multiple treatments.

The setting of multiple treatment options is important in general and especially so in an era of personalized or precision medicine. Full discussion of the issue is beyond the scope of the present paper, but many of the points discussed above do generalize to the multiple treatments setting. Specifically, in the task of optimizing treatment decisions without resource constraints, the solution to maximizing the population outcome, which is itself identical, in this setting, to maximizing the outcome for each individual involves a very similar form to what has already been discussed above. The optimal treatment rule in this setting with measured covariates C simply involves obtaining the expected outcome given an individual's covariates C under each possible treatment, $E[Y|A=a,C=c]$, $a=0,1,2,…,N$, and then assigning to each individual the treatment that gives the highest predicted outcome. Likewise, for the same reasons as those given above, in this setting, if the goal is to maximize population outcomes or individual decision-making, there is little reason to carry out one-by-one-covariate subgroup analyses or to consider which product terms in statistical models are statistically significant.

Where then does this leave us? If we knew the exact form of the expected outcomes conditional on covariates $E[Y|A=a,C=c]$ separately for each treatment group, the problem would essentially be solved. We simply assign the treatment that gives the highest predicted outcome. The task simply becomes obtaining the most predictive covariates C and obtaining the predictions. In actual fact, the functional form of the conditional expected outcome is unknown and must be estimated and this becomes

a difficult statistical task. In this paper, we have pointed out some statistical methodology and software to carry this out in some settings. However, again, important work remains to be done in determining at what sample sizes, and for how many covariates, and for how many treatment levels this is in fact feasible.

In the clinical setting, one might also wonder about the role of expert judgment. Are there perhaps aspects of a patient's profile that are not, or even cannot be, adequately captured by a variable that we can use in a statistical model? This of course remains a possibility. Are we to abandon clinical judgment and simply rely on statistical models to make such predictions? Are we to pit clinical judgment and modeling against one another? We would like to close this paper by attempting to tackle this question head on with a compromise, to allow both clinical judgment and predictive models, by proposing a new type of study design.

A possible design – what we will refer as an "Expected Outcomes Trial" – is to first use either *prior* randomized trial data, and/or observational data, with a relatively rich set of covariates C to build models for the expected outcomes with and without treatment. With such models, for each study participant in the Expected Outcomes Trial, the clinician (or patient) is randomized either to receive no further information, or to receive information on the expected outcome given their covariates under each treatment scenario. This could include outcomes under multiple treatment options. The clinician (or patient) then decides, based on the information available and their own judgments and preferences, which treatment to select. Outcomes are measured after a suitable follow-up period to determine whether the information provided by the predictive outcome models is useful in such decision-making. A trial of this sort will allow decision-makers to make use of both individually-oriented outcome predictions under statistical models, and also personal judgments, in making treatment decisions. It would also preserve decision-maker autonomy, and be more likely to be palatable to clinicians, and therefore more likely also to be scalable. The trials themselves would determine the additional utility of the information provided by the predictive models. A variation that added an additional arm in which treatment always followed the predicted maximum outcome could also be used to evaluate the role of clinical judgement, whether beneficial or harmful, above and beyond reliance on predicted probabilities.

We believe that such trials will be of use in determining the utility of prediction models for personalized or precision medicine in actual practical settings. Moreover, we believe that careful thought as to the what the correct question is in individual treatment decision-making, and careful selection of the correct optimization question and statistical method corresponding to the question of interest, will result in better patient outcomes. Current practices of one-covariate-at-a-time subgroup analysis, the use of prognostic scores, and the detection of significant interactions are simply not optimal for decision-making.

# eAppendix for "Selecting optimal subgroups for treatment using many covariates"

## A. The Form of the Optimal Treatment Rule

### Notation

Let $A$ denote a binary treatment of interest, $Y$ an outcome and $C$ a set of measured baseline covariates. Let $Y_1$ and $Y_0$ denote the potential outcomes for each individual under treatment levels 1 and 0 respectively. Let the population of individuals be denoted by $\Omega$. We first assume treatment $A$ is randomized and then consider treatment that may arise from an observational study. For simplicity in the next several section, in order to give intuitive proofs, we will assume a finite population of individuals with no ties at the cut-off for the optimal treatment rule. See Luedtke and van der Laan (2015, 2016ab) for further discussion of these cases without these conditions.

### Context 1: Treatment Subgroup Selection Under Limited Resources

Suppose that due to limited resources we can only treat $100q\%$ of the population. We will assume that treatment is beneficial for at least $100q\%$ of the population i.e. $P(Y_1 - Y_0 > 0) > q$. Otherwise, if we know this in advance, the problem reduces to Context 2 described below. We desire to partition $\Omega$ into sets $S$ and $T$ such that $P(\omega \in T) = q$ so as to maximize the average outcome if all units in $T$ were treated and all units in $S$ were untreated. In other words, we wish to identify $S$ and $T$ to solve:

$$\arg\max_{S,T:S \cup T = \Omega, S \cap T = \emptyset, P(\omega \in T) = q} [qE(Y_1|\omega \in T) + (1-q)E(Y_0|\omega \in S)].$$

In fact, choosing $S$ and $T$ to maximize the average outcome if all units in $T$ were treated and all units in $S$ were untreated is equivalent to choosing $S$ and $T$ to maximize the treatment effect heterogeneity with $100q\%$ in one group, as stated in the following proposition.

**Proposition 1.** *The solution to*

$$\arg\max_{S,T:S \cup T = \Omega, S \cap T = \emptyset, P(\omega \in T) = q} [qE(Y_1|\omega \in T) + (1-q)E(Y_0|\omega \in S)]$$

*is equivalent to the solution to*

$$\arg\max_{S,T:S \cup T = \Omega, S \cap T = \emptyset, P(\omega \in T) = q} [E(Y_1 - Y_0|\omega \in T) - E(Y_1 - Y_0|\omega \in S)].$$

*Proof.* We have that

$$\begin{aligned}
&\arg\max_{S,T:S \cup T = \Omega, S \cap T = \emptyset, P(\omega \in T) = q} [E(Y_1 - Y_0|\omega \in T) - E(Y_1 - Y_0|\omega \in S)] \\
=\ &\arg\max_{S,T:S \cup T = \Omega, S \cap T = \emptyset, P(\omega \in T) = q} [E(Y_1|\omega \in T) - E(Y_1|\omega \in S) - E(Y_0|\omega \in T) + E(Y_0|\omega \in S) \\
&\quad -\frac{1}{1-q}E(Y_1) + \frac{1}{q}E(Y_0) \\
=\ &\arg\max_{S,T:S \cup T = \Omega, S \cap T = \emptyset, P(\omega \in T) = q} [E(Y_1|\omega \in T) - E(Y_1|\omega \in S) - E(Y_0|\omega \in T) + E(Y_0|\omega \in S) \\
&\quad -\frac{q}{1-q}E(Y_1|\omega \in T) - \frac{1-q}{1-q}E(Y_1|\omega \in S) + \frac{q}{q}E(Y_0|\omega \in T) + \frac{(1-q)}{q}E(Y_0|\omega \in S)] \\
=\ &\arg\max_{S,T:S \cup T = \Omega, S \cap T = \emptyset, P(\omega \in T) = q} [\frac{1-2q}{1-q}E(Y_1|\omega \in T) + \frac{1-2q}{q}E(Y_0|\omega \in S)] \\
=\ &\arg\max_{S,T:S \cup T = \Omega, S \cap T = \emptyset, P(\omega \in T) = q} [qE(Y_1|\omega \in T) + (1-q)E(Y_0|\omega \in S)].
\end{aligned}$$

□

The solution to this maximization problem in fact takes a very simple form as stated in the next proposition.

**Proposition 2.** *The solution to*

$$\arg\max_{S,T:S\cup T=\Omega, S\cap T=\emptyset, P(\omega\in T)=q}[E(Y_1 - Y_0|\omega \in T) - E(Y_1 - Y_0|\omega \in S)]$$

*almost surely takes the form, for some $\kappa$, of $T = \{\omega \in \Omega : Y_1(\omega) - Y_0(\omega) > \kappa\}$ and $S = \{\omega \in \Omega : Y_1(\omega) - Y_0(\omega) \le \kappa\}$.*

*Proof.* We prove the result by contradiction. Suppose $S$ and $T$ were not of this form. Then there must exist (possibly non-unique) disjoint sets $\Omega' \subseteq T$ and $\Omega^* \subseteq S$ of equal and positive probability such that $Y_1(\omega') - Y_0(\omega') < Y_1(\omega^*) - Y_0(\omega^*)$ for all $\omega' \in \Omega'$ and $\omega^* \in \Omega^*$. Let $T' = (T \cup \Omega^*) \setminus \Omega'$ and $S' = (S \cup \Omega') \setminus \Omega^*$. Then $E(Y_1 - Y_0|\omega \in T') - E(Y_1 - Y_0|\omega \in S') > E(Y_1 - Y_0|\omega \in T) - E(Y_1 - Y_0|\omega \in S)$ and thus $S$ and $T$ would not be the solution to $\arg\max_{S,T:S\cup T=\Omega, S\cap T=\emptyset, P(\omega\in T)=q}[E(Y_1 - Y_0|\omega \in T) - E(Y_1 - Y_0|\omega \in S)]$. $\square$

If both counterfactual outcomes were known for all individuals, then $\kappa$ could be obtained as the solution to $P(Y_1 - Y_0 > \kappa) = q$ and the average outcome for the population under the optimal treatment rule of giving treatment if $Y_1 - Y_0 > \kappa$ would be $qE(Y_1|Y_1 - Y_0 > \kappa) + (1-q)E(Y_0|Y_1 - Y_0 \le \kappa)$. It is of course not possible to partition individuals in this way without complete knowledge of the counterfactual outcomes, which will in general not be available.

However, it is still possible to partition the covariate space to carry out a similar maximization. If we let $\Gamma$ denote the support of $C$ the task becomes

$$\arg\max_{S,T:S\cup T=\Gamma, S\cap T=\emptyset, P(C\in T)=q}[E(Y_1 - Y_0|\omega \in T) - E(Y_1 - Y_0|\omega \in S)].$$

By arguments similar to those presented above, the solution to this takes the form of $T = \{c \in \Gamma : E(Y|A=1,c) - E(Y|A=0,c) > k\}$ i.e. the treatment rule is then simply give treatment to those for whom $E(Y|A=1,c) - E(Y|A=0,c) > k$ with $k$ given as the solution to

$$\int 1[E(Y_1 - Y_0|C=c) > k]dP(c) = q.$$

In practice one must model $E(Y|A=1, C=c) - E(Y|A=0, C=c)$ and estimate $k$.

The average outcome under this treatment rule is then given by $q\int E(Y|A=1,c)dP(c|C \in T) + (1-q)\int E(Y|A=0,c)dP(c|C \notin T)$. Note that this average outcome will not in general be as high as the average outcome under the optimal decision rule if both counterfactual outcomes were themselves known for all individuals, in which case the average outcome would be given as above, $qE(Y_1|Y_1 - Y_0 > \kappa) + (1-q)E(Y_0|Y_1 - Y_0 \le \kappa)$ where $\kappa$ is defined as the solution to $P(Y_1 - Y_0 > \kappa) = q$. The extent to which the average outcome under the optimal treatment rule using the measured covariates $C$ comes close to that which could be obtained under complete knowledge of the counterfactual outcomes will depend on the extent to which the covariates $C$ are predictive of the outcome itself. Note that it is also thus of course the case that the average outcome under the optimal treatment rule using data on the measured covariates $C$ will always be relative to $C$.

It may be of interest to compare the average outcome under this optimal treatment decision rule using the measured covariates $C$ to the average outcome with no one treated, $E(Y|A=0)$, the outcome with $100q\%$ treated but selected randomly $qE(Y|A=1) + (1-q)E(Y|A=0)$, and the average outcome with everyone treated, $E(Y|A=1)$. It might also be of interest to compare the treatment effect for those treated under the optimal rule, $\int \{E(Y|A=1,c) - E(Y|A=0,c)\}dP(c|C \in T)$, to those left untreated by the rule $\int \{E(Y|A=1,c) - E(Y|A=0,c)\}dP(c|C \notin T)$, and also to the average treatment effect for the population $E(Y|A=1) - E(Y|A=0)$. Another relevant metric may be taken as the difference between the treatment effects comparing those who are assigned treatment by the rule versus those who are not:

$$\int \{E(Y|A=1,c) - E(Y|A=0,c)\}dP(c|C \in T)$$

$$-\int \{E(Y|A=1,c) - E(Y|A=0,c)\}dP(c|C \notin T).$$

This could be taken as a measure of effect heterogeneity introduced by the optimal treatment rule for the $100q\%$ to receive treatment. The analogous metric under treatment of a random $100q\%$ of the population selected for treatment would simply be 0.

## Context 2: Unconstrained Treatment Subgroup Selection

Now suppose that resources are unconstrained and all could be treated who benefit from treatment. The optimal subgroup for treatment would then be

$$T = \{\omega : Y_1(\omega) - Y_0(\omega) > 0\}.$$

Once again, we cannot determine this subgroup as in general we will not have information on both potential outcomes for all individuals. Instead, with covariate data on $C$, we would select the subgroup to treat as those with covariates $C$ such that the expected value of the treatment effect conditional on $C$ was positive. Thus we would treat those with covariate values that lie in the set

$$T = \{c \in \Gamma : E(Y|A=1,c) - E(Y|A=0,c) > 0\}.$$

In practice with high dimensional $C$, the expectation $E(Y|a,c)$ would have to be modeled. If we were to follow this treatment rule, the average outcome under this rule would be

$$\int_{c \in T} E(Y|A=1,c)dP(c) + \int_{c \notin T} E(Y|A=0,c)dP(c).$$

Note that this average outcome will not in general be as high as the average outcome under the optimal decision rule if both counterfactual outcomes were themselves known in which case the average outcome would be $E(Y_1|Y_1 - Y_0 > 0)P(Y_1 - Y_0 > 0) + E(Y_0|Y_1 - Y_0 \leq 0)P(Y_1 - Y_0 \leq 0)$. The extent to which the average outcome under the optimal treatment rule using the measured covariates $C$ comes close to that which could be obtained under complete knowledge of the counterfactual outcomes will depend on the extent to which the covariates $C$ are predictive of the outcome itself. It is also thus of course the case that the average outcome under the optimal treatment rule using data on the measured covariates $C$ will always be relative to $C$.

It may be of interest to compare this average outcome under the optimal treatment rule using the measured covariates $C$ to the average outcome with no one treated, $E(Y|A=0)$, and to the average outcome with everyone treated, $E(Y|A=1)$. It might also be of interest to compare the treatment effect for those treated under the optimal rule $\int \{E(Y|A=1,c) - E(Y|A=0,c)\}dP(c|C \in T)$, to those left untreated by the rule $\int \{E(Y|A=1,c) - E(Y|A=0,c)\}dP(c|C \notin T)$, and also to the average treatment effect for the population $E(Y|A=1) - E(Y|A=0)$. Another relevant metric may be taken as the difference between the treatment effects comparing those who are assigned treatment by the rule versus those who are not:

$$\int \{E(Y|A=1,c) - E(Y|A=0,c)\}dP(c|C \in T) - \int \{E(Y|A=1,c) - E(Y|A=0,c)\}dP(c|C \notin T).$$

This could be taken as a measure of effect heterogeneity.

## Context 3: Unconstrained Treatment Subgroup Selection Under Costs or Side Effects

Another generalization of Setting 1 that might be considered is a setting in which there is a cost constraint. Typically this will require the investigator to only treat those with treatment effect greater than some threshold $\delta(c)$ that relies on the covariate value $c$. If all counterfactuals were known, we would use individual-level treatment effect $Y_1(\omega) - Y_0(\omega)$, and we will show that this can be replaced by $E(Y|A=1,c) - E(Y|A=0,c)$ in the realistic setting where only one counterfactual is observed for each individual.

To formalize our discussion, we wish to estimate the solution to

$$\text{Maximize} \int_{c \in T} E(Y|A=1,c)dP(c) + \int_{c \notin T} E(Y|A=0,c)dP(c)$$

$$\text{subject to} \int_{c \in T} \text{Cost}(c)dP(c) \leq \text{Cost Constraint},$$

where $\text{Cost}(\cdot)$ is pre-defined positive function giving the cost of treating someone in covariate strata and Cost Constraint is the pre-defined constraint. The results in this section easily generalize to the case where $\text{Cost}(c)$ can equal zero, but we omit this case for simplicity. If $b = E[\text{Cost}(C)]$ is less than Cost Constraint then the constraint is not active and we revert to Context 2. Otherwise the $T$ maximizing the above objective takes the following form.

**Proposition 3.** *If $b$ is finite and greater than* Cost Constraint, *then the $T$ maximizing the objective function above takes the form, $\{\omega \in \Omega : E(Y|A=0,c) - E(Y|A=0,c) > k\,\text{Cost}(c)\}$, where $k$ is the solution to*

$$\int 1\left[E(Y_1 - Y_0|C=c) > k\,\text{Cost}(c)\right]\text{Cost}(c)dP(c) = \text{Cost Constraint}.$$

*Proof.* Let $P_c$ denote the probability measure with density $\frac{dP_c}{dP}(c) = \text{Cost}(c)/b$. Let $\tilde{Y} = bY/\text{Cost}(C)$, where we note that $\tilde{Y}$ is a deterministic function of $Y$ conditional on $C = c$ so that $bE_P(Y|A=a,c)/\text{Cost}(c)$ is equal to $E_P(\tilde{Y}|A=a,c)$. We can write the mean outcome under the optimal rule as

$$\int_{c \in T} E(\tilde{Y}|A=1,c)dP_c(c) + \int_{c \notin T} E(\tilde{Y}|A=0,c)dP_c(c).$$

The cost constraint rewrites as $P_c(C \in T) \leq (\text{Cost Constraint})/b$. We are now in Context 1, so maximizing $T$ takes the form $\{\omega \in \Omega : E(\tilde{Y}|A=0,c) - E(\tilde{Y}|A=0,c) > \tilde{k}\}$, where $\tilde{k}$ is the solution to

$$\int 1\left[E(\tilde{Y}|A=1,c) - E(\tilde{Y}|A=0,c) > \tilde{k}\right]dP_c(c) = (\text{Cost Constraint})/b$$

Multiplying both sides by $b$, using the definition of $\tilde{Y}$, and letting $k = \tilde{k}/b$ gives the result. □

It may be of interest to compare this average outcome under the optimal treatment rule to the average outcome with no one treated, $E(Y|A=0)$, and to the average outcome with everyone treated, $E(Y|A=1)$. It might also be of interest to compare the treatment effect for those treated under the optimal rule, $\int \{E(Y|A=1,c) - E(Y|A=0,c)\}dP(c|C \in T)$, to those left untreated by the rule $\int \{E(Y|A=1,c) - E(Y|A=0,c)\}dP(c|C \notin T)$, and also to the average treatment effect for the population $E(Y|A=1) - E(Y|A=0)$.

## Context 4: Treatment Subgroup Selection to Maximize Treatment Effect Heterogeneity

In examining subgroup analyses reported in the literature one is sometimes under the impression that a central goal is to find a subgroup division that maximizes effect heterogeneity across the subgroups.

While it is not clear that this goal is of principal policy importance, it can be carried out using a set of measured covariates $C$. The task could then be stated as finding a partition of individuals into sets $T$ and $S$ to maximize treatment effect heterogeneity when comparing the treatment effects among those in $T$ versus $S$. The problem could thus formally be stated as

$$\arg\max_{S,T:S\cup T=\Omega, S\cap T=\emptyset}[E(Y_1 - Y_0|\omega \in T) - E(Y_1 - Y_0|\omega \in S)].$$

The solution to this maximization problems takes a relatively simple form as in the next Proposition.

**Proposition 4.** *The solution to*

$$\arg\max_{S,T:S\cup T=\Omega, S\cap T=\emptyset}[E(Y_1 - Y_0|\omega \in T) - E(Y_1 - Y_0|\omega \in S)]$$

*takes the form, for some $\kappa$, of $T = \{\omega \in \Omega : Y_1(\omega) - Y_0(\omega) > \kappa\}$ and $S = \{\omega \in \Omega : Y_1(\omega) - Y_0(\omega) \leq \kappa\}$.*

*Proof.* The proof is similar to that of Proposition 2 above. □

If we let $V = Y_1 - Y_0$ and let $p(v)$ denote the density of $V$. To determine $\kappa$, we wish to choose $\kappa$ to maximize $E(Y_1 - Y_0|Y_1 - Y_0 > \kappa) - E(Y_1 - Y_0|Y_1 - Y_0 \leq \kappa)$ or equivalently,

$$\int_\kappa^\infty vp(v)dv / \int_\kappa^\infty p(v)dv - \int_{-\infty}^\kappa vp(v)dv / \int_{-\infty}^\kappa p(v)dv.$$

Differentiating with respect to $\kappa$ we obtain:

$$\frac{\int_\kappa^\infty p(v)dv [vp(v)]_{v=\kappa}^{v=\infty} - \int_\kappa^\infty vp(v)dv [p(v)]_{v=\kappa}^{v=\infty}}{[\int_\kappa^\infty p(v)dv]^2} - \frac{\int_{-\infty}^\kappa p(v)dv [vp(v)]_{v=-\infty}^{v=\kappa} - \int_{-\infty}^\kappa vp(v)dv [p(v)]_{v=-\infty}^{v=\kappa}}{[\int_{-\infty}^\kappa p(v)dv]^2}$$

Setting this equal to 0 and solving for $\kappa$ gives

$$[\int_{-\infty}^\kappa p(v)dv]^2 \{\int_\kappa^\infty p(v)dv [vp(v)]_{v=\kappa}^{v=\infty} - \int_\kappa^\infty vp(v)dv [p(v)]_{v=\kappa}^{v=\infty}\} - [\int_\kappa^\infty p(v)dv]^2 \{\int_{-\infty}^\kappa p(v)dv [vp(v)]_{v=-\infty}^{v=\kappa}$$

$$- \int_{-\infty}^\kappa vp(v)dv [p(v)]_{v=-\infty}^{v=\kappa}\} = 0$$

$$P(V \leq \kappa)^2 \quad \{-\kappa p(\kappa)P(V > \kappa) + p(\kappa)\int_\kappa^\infty vp(v)dv\} + P(V > \kappa)^2 \quad \{-\kappa p(\kappa)P(V \leq \kappa) + p(\kappa)\int_{-\infty}^\kappa vp(v)dv\} = 0$$

$$P(V \leq \kappa)^2 \quad \{-\kappa P(V > \kappa) + \int_\kappa^\infty vp(v)dv\} + P(V > \kappa)^2 \quad \{-\kappa P(V \leq \kappa) + \int_{-\infty}^\kappa vp(v)dv\} = 0.$$

The solution to the equation gives $\kappa$ which could be obtained numerically.

The treatment rule that maximizes effect heterogeneity is then either $T = \{\omega \in \Omega : Y_1(\omega) - Y_0(\omega) \geq \kappa\}$ or $T = \{\omega \in \Omega : Y_1(\omega) - Y_0(\omega) > \kappa\}$. The maximum treatment effect heterogeneity that can thus be obtained comparing two subgroups that partition all individuals is thus $E(Y_1 - Y_0|Y_1 - Y_0 > \kappa) - E(Y_1 - Y_0|Y_1 - Y_0 \leq \kappa)$.

It is of course not possible to partition individuals in this manner without complete knowledge of the counterfactual outcomes. However, it is still possible to carry out a similar partitioning using measured covariates $C$. If we let $\Gamma$ denote the support of $C$, the task becomes

$$\arg\max_{S,T\subseteq\Gamma: S\cup T=\Gamma, S\cap T=\emptyset}[E(Y_1 - Y_0|C \in T) - E(Y_1 - Y_0|C \in S)].$$

By arguments similar to those presented above, the sets $S$ and $T$ are then given by $T = \{c \in \Gamma : E(Y|A = 1, c) - E(Y|A = 0, c) > k\}$ and $S = \{c \in \Gamma : E(Y|A = 1, c) - E(Y|A = 0, c) \leq k\}$ for some $k$, and once again $k$ could be solved for numerically.

The effect heterogeneity under this treatment rule for maximizing effect heterogeneity with measured covariates $C$ is then given by:

$$\int \{E(Y|A=1,c) - E(Y|A=0,c)\}dP(c|C \in T) - \int \{E(Y|A=1,c) - E(Y|A=0,c)\}dP(c|C \notin T).$$

Note that this measure of effect heterogeneity will not in general be as high as the maximum effect heterogeneity if both counterfactual outcomes for all individuals were known, which was given above as $E(Y_1 - Y_0|Y_1 - Y_0 \geq \kappa) - E(Y_1 - Y_0|Y_1 - Y_0 < \kappa)$. The extent to which the maximum effect heterogeneity under the treatment rule using the measured covariates $C$ comes close to the maximum which can be obtained under complete knowledge of the counterfactual outcomes will depend on the extent to which the covariates $C$ are predictive of the outcome itself. Note that it is also thus of course the case that the maximum effect heterogeneity under the treatment rule using data on the measured covariates $C$ will always be relative to $C$. It might also be of interest to compare the treatment effect for those treated under the maximizing rule, $\int \{E(Y|A=1,c) - E(Y|A=0,c)\}dP(c|C \in T)$, to those left untreated by the rule $\int \{E(Y|A=1,c) - E(Y|A=0,c)\}dP(c|C \notin T)$, and also to the average treatment effect for the population $E(Y|A=1) - E(Y|A=0)$.

If using the treatment rule that maximizes effect heterogeneity, one can also estimate the average outcome under this rule as

$$\int_{c \in T} E(Y|A=1,c)dP(c) + \int_{c \notin T} E(Y|A=0,c)dP(c).$$

One could compare this to the average outcome with no one treated, $E(Y|A=0)$, and the average outcome with everyone treated, $E(Y|A=1)$. Note, however, this average outcome under the treatment rule that maximizes effect heterogeneity will in general be lower than the average outcome under the treatment rule that maximizes the average outcome itself as discussed in Context 2. It is thus not clear that this treatment rule that maximizes effect heterogeneity is of particular use in policy-making, unlike contexts 1, 2 and 3 above.

### Further Comments on Observational Studies

The above approaches and results for randomized treatment $A$ apply also to observational studies in which the covariates $C$ suffice to control for confounding of the effect of $A$ on $Y$ but, when reference is made to the average outcome when no one is treated, $E(Y|A=0)$ must be replaced by $\int E(Y|A=0,c)dP(c)$; when reference is made to the average outcome when everyone is treated, $E(Y|A=1)$ must be replaced by $\int E(Y|A=1,c)dP(c)$; and when reference is made to the average treatment effect for the population $E(Y|A=1) - E(Y|A=0)$ must be replaced by $\int E(Y|A=1,c)dP(c) - \int E(Y|A=0,c)dP(c)$.

## B. Statistical Estimation

Supose we observe an i.i.d. sample $(C_1, A_1, Y_1), \ldots, (C_n, A_n, Y_n)$ of observations generated according to a randomized trial. We now consider estimation of the parameter from Context 2 in this setting. All of these estimation problems rely on the user having an estimate of the conditional average treatment effect (CATE), i.e. the function mapping rom a covariate $c$ to $E[Y|A = 1, c] - E[Y|A = 0, c]$. We therefore start by presenting an estimation procedure for this quantity, and then proceed to give an overview of estimation for Contexts 1 and 2. A specific estimation procedure for Contexts 1 and 2 can be found in the eAppendix.

### Estimation of the Conditional Average Treatment Effect

We frame the estimation of the CATE as a regression problem, which thereby enables both the use of classical linear regression approaches and also any machine learning algorithm which is designed to estimate a conditional mean function.

For any triplet of observations $(c, a, y)$ and a given function $f$ mapping from $(a, c)$ to the real line, define the pseudo-outcome

$$\tilde{y} = \frac{2a - 1}{P(A = a|C = c)}[y - f(a, c)] + f(1, c) - f(0, c),$$

where we recall that the probability of treatment given covariates is known in our randomized trial setting. We use $\tilde{Y}_1, \ldots, \tilde{Y}_n$ to denote the corresponding random variable for our trial participants and $\tilde{Y}$ to denote the general random variable corresponding to $(C, A, Y)$. We will provide guidance on the selection of $f$ at the end of this section. Observe that

$$E\left[\tilde{Y}\,\middle|\, C = c\right] = E[Y|A = 1, c] - E[Y|A = 0, c].$$

This justifies the following estimation procedure for the CATE function:

1. Define pseudo-observations $\tilde{Y}_1, \ldots, \tilde{Y}_n$.

2. Regress $(\tilde{Y}_1, \ldots, \tilde{Y}_n)$ against $(C_1, \ldots, C_n)$ using a preferred regression algorithm.

Suppose for simplicity than one restricts oneself to least squares regression techniques for the latter step (possibly subject to some complexity penalty). The class of such techniques is far richer than the classical linear regression techniques: for example, generalized additive models, kernel smoothers, and smoothing splines can all be fit using this criterion (Hastie et al., 2002). The estimate for the CATE resulting from the above will typically provide a mean-square consistent estimate if the regression algorithm is correctly specified, and otherwise one can still measure the quality of a given estimate using mean-squared error (MSE). Furthermore, if one has two correctly specified regression algorithms so that both have MSE converging to zero with sample size, then typically the MSE of the "simpler" estimator decays at a faster rate. As a simple example, a correctly specified univariate linear regression will typically have smaller MSE than a correctly specified random forest regression in finite samples.

Committing to a single regression algorithm in the second step above is therefore unadvisable: ideally one wants to select the simplest estimator which is (nearly) correctly specified so that the MSE is small for the finite sample, but doing so *a priori* is not generally possible. To overcome this difficulty, we propose the use of the ensemble algorithm known as super-learning (van der Laan, 2007). This algorithm allows the user to input a library of candidate regression algorithms and outputs the fit from a convex combination of the candidate algorithms. This convex combination is selected to minimize the cross-validated MSE between $\tilde{Y}$ and the fit. The super-learner algorithm is optimal in the sense that the cross-validated MSE of the resulting fit is at least as good as the cross-validated MSE of the best algorithm in the library (up to a small remainder term). The usefulness of this finite sample result has been repeatedly supported by simulations in the literature (cf. Polley and van der Laan, 2010; van der Laan and Rose, 2011). See Luedtke and van der Laan (2014) for theoretical foundations and simulation results specific to the CATE.

We conclude by discussing the selection of the function $f$. As the described method is valid for any fixed function $f$, one can set $f$ equal to the constant function zero so that the pseudo-outcome takes on a simple form. Though appealing for its simplicity, this choice of $f$ will generally yield a suboptimal fit of the CATE. Given that it is typically easier to fit a regression when the conditional variance of the outcome given covariates is smaller, one may wish to choose $f$ to make this quantity small. The choice of $f$ which minimizes this quantity is given by $f(a,c) = E[Y|a,c]$. In practice $E[Y|a,c]$ is not known, but typically substituting an estimate $\hat{E}[Y|a,c]$ will perform better than choosing $f$ equal to zero (see Luedtke and van der Laan (2014) for a deeper discussion, and a double robustness based justification for selecting $f$ in this way that will prove useful in observational studies).

## Estimates for Contexts 1 and 2

In the eAppendix we provide estimates for the parameters arising in Contexts 1 and 2 given an i.i.d. sample $(C_1, A_1, Y_1), \ldots, (C_n, A_n, Y_n)$ generated from a randomized trial. We propose using the cross-validated targeted minimum loss-based estimation (CV-TMLE) framework to estimate these quantities. This framework allows the user to provide initial estimates of the outcome regression $E[Y|a,c]$ and the CATE, where we do not require that the CATE estimate is equal to the difference of outcome regressions at $A = 1$ and $A = 0$. We advise using the super-learner methodology for the CATE discussed in the previous section, and a standard regression-based super-learner for the outcome regression, regressing $Y$ against treatment and covariates.

At this point one might be tempted to use as estimate the substitution estimator

$$\frac{1}{n} \sum_{i=1}^{n} 1\left(\widehat{\mathrm{CATE}}(C_i) > \delta_n\right) \left(\hat{E}[Y|1, C_i] - \hat{E}[Y|0, C_i]\right),$$

where $\delta_n = 0$ for Context 1 and $\delta_n$ is the cutoff estimated from the data for Context 2. Unfortunately, this estimate will tend to be too biased for the parameter of interest, thereby making root-$n$ rate inference impossible. For intuition we will first discuss the non-cross-validated TMLE procedure, and then discuss how the CV-TMLE procedure differs. The TMLE procedure confronts this challenge by fluctuating the initial outcome regression estimate $\hat{E}[Y|a,c]$ with a univariate submodel selected to reduce the bias of the above substitution estimator for the parameter of interest. Though a thorough description of the selected submodel is beyond the scope of this manuscript, we invite readers to explore Luedtke and van der Laan (2015) and van der Laan and Luedtke (2014) for the justification of the submodels used in Contexts 1 and 2, respectively. Given a fluctuated outcome regression estimate $\hat{E}^*[Y|a,c]$, one can then use the less biased substitution estimator

$$\frac{1}{n} \sum_{i=1}^{n} 1\left(\widehat{\mathrm{CATE}}(C_i) > \delta_n\right) \left(\hat{E}^*[Y|1, C_i] - \hat{E}^*[Y|0, C_i]\right)$$

Though one can establish regularity conditions under which the bias is asymptotically negligible, it is difficult to establish reasonable conditions exerting control over the finite sample bias of this estimator, especially in Context 2 where we are attempting to make inference on a quantity that the true CATE maximizes over. Though we may use a different criteria function to estimate the CATE, this estimator will still be prone to positive bias in finite samples. This positive bias manifests itself similarly to the negative bias that results when one attempts to estimate the squared error between an outcome $Y$ and estimated regression function $\hat{E}[Y|X]$ using the empirical squared error.

A natural solution to this finite sample bias issue is to use cross-validation. In the appendix we present a cross-validated form of the TMLE algorithm for which this bias should not appear. Under a mild consistency condition, our final estimator $\hat{\psi}$ of the parameters $\psi$ from Context 1 satisfies the identity

$$\hat{\psi} - \psi \approx \frac{1}{n} \sum_{i=1}^{n} D_0(O_i) + \frac{1}{10} \sum_{v=1}^{10} E\left[\left(I(\widehat{CATE}_v(C) > 0) - I(E[Y|1,C] - E[Y|0,C] > 0)\right)(E[Y|1,C] - E[Y|0,C])\right]$$

for a fixed function $D_0$, where each $\widehat{CATE}_v$ is an estimate of the CATE fit on 9/10 of the data and the approximation holds up to a remainder term that shrinks to zero faster than $n^{-1/2}$ in probability. If the second term on the right is negligible, then we can develop standard Wald-type confidence intervals for the parameter of interest using the central limit theorem. Theorem 8 in Luedtke and van der Laan (2016) shows that the second term on the right is indeed sufficiently small under a variety of assumptions on the rate of convergence of the estimate of the CATE, one of which is a condition on its MSE. A similar expansion holds for Context 2, though we omit it for brevity (see Luedtke and van der Laan (2015) for details). Finally, we observe that the second term on the right-hand side above is nonpositive. Hence, even if this term fails to be sufficiently small to be asymptotically negligible, our estimator for the parameter of interest will at worst be negatively biased. One can further show that this implies to establish the asymptotic validity of the lower bound of our confidence interval regardless of the quality of our estimate of the CATE.

### Super-Learner for the Conditional Average Treatment Effect

The super-learner algorithm for this estimation problem is given below. For simplicity we use ten-fold cross-validation and assume that the sample size $n$ is a multiple of ten. We suppose that the user has $m$ candidate regression algorithms.

1. Define pseudo-observations $\tilde{Y}_1, \ldots, \tilde{Y}_n$.

2. For $v$ in $\{1, \ldots, 10\}$:

   (a) For $\ell$ in $\{1, \ldots, m\}$:
   
   - Fit candidate algorithm $\ell$ on observations $\left((C_i, \tilde{Y}_i) : i \notin \left\{\frac{(v-1)n}{10} + 1, \ldots, \frac{vn}{10}\right\}\right)$ to generate the regression function estimate that takes as input $c$ and outputs $\hat{E}_v^\ell[\tilde{Y}|c]$.
   
   (b) For $i$ in $\left\{\frac{(v-1)n}{10} + 1, \ldots, \frac{vn}{10}\right\}$, define $X_i^{\text{CV}}$ to be the $m$-length column vector $\left(\hat{E}_v^\ell[\tilde{Y}|c_i] : \ell = 1, \ldots, m\right)$.

3. Choose $\alpha_n$ be the convex $m$-length row vector minimizing $\frac{1}{n}\sum_{i=1}^n \left(\tilde{Y}_i - \alpha_n X_i^{\text{CV}}\right)^2$.

4. For $\ell$ in $\{1, \ldots, m\}$:

   - Fit the candidates $\ell$ on all of the observations, yielding a function which takes as input $c$ and outputs $\hat{E}^\ell[\tilde{Y}|c]$.

5. Return the function $\hat{b}$ defined by $\hat{b}(c) = \alpha_n X(c)$, where $X(c)$ is the column vector $\left(\hat{E}^\ell[\tilde{Y}|c] : \ell = 1, \ldots, m\right)$.

### Estimator for Contexts 1 and 2

We now present a cross-validated targeted minimum loss-based estimator (CV-TMLE) for Contexts 1 and 2. We omit the derivation of this estimator from this work for brevity, and instead refer the reader to references given in the main text for the derivation of CV-TMLEs for a nearly identical parameters as in Contexts 1 and 2. We assume that the outcome $Y$ is bounded, and without loss of generality we assume that the lower bound is 0 and the upper bound is 1.

For simplicity, we use ten-fold cross-validation and assume that the sample size $n$ is a multiple of ten. For each $v = 1, \ldots, 10$, we let $T_v$ denote the observations $(C_i, A_i, Y_i)$ with indices $i \notin \left\{\frac{(v-1)n}{10} + 1, \ldots, \frac{vn}{10}\right\}$, i.e. the observations included in the training sample for a given cross-validation split. We also let $v(i)$ denote the $v$ such that $i \notin \left\{\frac{(v-1)n}{10} + 1, \ldots, \frac{vn}{10}\right\}$, i.e. the cross-validation index for which $i$ is not in the training sample.

1. For $v$ in $\{1, \ldots, 10\}$:

(a) Obtain an estimate $\hat{E}_v[Y|\cdot]$ of the function mapping from $(a,c)$ to $E[Y|a,c]$ using observations in $T_v$.
   (b) Obtain an estimate $\hat{b}_v$ of the CATE function using observations in $T_v$.
   Note: $\hat{b}_v(c)$ need not equal $\hat{E}[Y|A=1,c] - \hat{E}[Y|A=0,c]$.

2. Let $\delta_n = 0$ in Context 2, and in Context 1 let $\delta_n$ be the positive part of the smallest solution in $\delta$ to
$$\frac{1}{n}\sum_{i=1}^n 1(\hat{b}_{v(i)}(c) > \delta) \le q.$$

3. Let $\epsilon_n$ be the slope estimate in an intercept-free logistic regression with outcome $(Y_i : i = 1,\ldots,n)$, covariate $\left(\frac{2A_i-1}{P(A_i|C_i)} : i = 1,\ldots,n\right)$, offset $\left(\text{logit}\,\hat{E}_{v(i)}[Y|A_i,C_i] : i = 1,\ldots,n\right)$, and observation weights $\left(1\left\{\hat{b}_{v(i)}(C_i) > \delta_n\right\} : i = 1,\ldots,n\right)$.
   Note: The `glm` function in R will run a logistic regression on any outcome bounded in $[0,1]$.

4. For each $v$ in $\{1,\ldots,10\}$ and any $(a,c)$, let $\bar{Q}^*_v(a,c) = \text{logit}^{-1}\left[\text{logit}\left(\hat{E}_v[Y|a,c]\right) + \epsilon_n \frac{2a-1}{P(a|c)}\right]$ represent the fluctuated estimate of $E[Y|a,c]$.

5. Let $\hat{\psi} = \frac{1}{n}\sum_{i=1}^n 1(\hat{b}_{v(i)}(C_i) > \delta_n)\left[\bar{Q}^*_{v(i)}(1,C_i) - \bar{Q}^*_{v(i)}(0,C_i)\right]$.

6. Let $D_i = \frac{2A_i-1}{P(A_i|C_i)}\left[Y_i - \bar{Q}^*_{v(i)}(A_i,C_i)\right] + \bar{Q}^*_{v(i)}(1,C_i) - \bar{Q}^*_{v(i)}(0,C_i) - \hat{\psi}$.

7. Let
$$\hat{\sigma}^2 = \frac{1}{n}\sum_{i=1}^n \left[1(\hat{b}_{v(i)}(C_i) > \delta_n)\left[D_i - \delta_n\right] + \delta_n q\right]^2,$$
   where we take $q = 1$ in Context 2.

8. Return the point estimate $\hat{\psi}$ and 95% confidence interval $\hat{\psi} \pm 1.96\frac{\hat{\sigma}}{\sqrt{n}}$.

## C. Software

To be completed.

# Additional References